\documentclass[%
reprint,
superscriptaddress,
amsmath,amssymb,
aps,longbibliography,
prl,
floatfix,
]{revtex4-1}

\usepackage{graphicx}%
\usepackage{verbatim}
\usepackage[version=4]{mhchem}
\usepackage{xcolor}
\usepackage{physics}
\usepackage[normalem]{ulem}
\usepackage{pdfpages}
\usepackage{pgffor}
\usepackage{xspace}

\usepackage[shortcuts]{glossaries}
\glsdisablehyper
\setacronymstyle{long-short}
\newacronym{dft}{DFT}{density-functional theory}
\newacronym{md}{MD}{molecular dynamics}
\newacronym{aimd}{aiMD}{\emph{ab initio} molecular dynamics}
\newacronym{pes}{PES}{potential-energy surface}
\newacronym{rmse}{RMSE}{root mean squared error}
\newacronym{rmsd}{RMSD}{root mean squared distance}
\newacronym{ad}{AD}{automatic differentiation}
\newacronym{ml}{ML}{machine learning}
\newacronym{pet}{PET}{Point Edge Transformer}
\newacronym[plural=MLPs,
	firstplural=machine-learning interatomic potentials]
	{mlp}{MLP}{machine-learning interatomic potential}

\newcommand{\mlp}{\gls{mlp}\xspace}
\newcommand{\mlps}{\glspl{mlp}\xspace}

\newcommand{\md}{\gls{md}\xspace}

\newcommand{\pet}{\gls{pet}\xspace}

\newcommand{\sok}{\textsc{So3krates}\xspace}
\newcommand{\nequip}{\textsc{Nequip}\xspace}
\newcommand{\mace}{\textsc{Mace}\xspace}

\newcommand{\SM}{Supp. Mat.\xspace}

\newcommand{\scare}[1]{\lq#1\rq}

\newcommand{\inv}{\ensuremath{\mathrm{i}}}

\usepackage[separate-uncertainty=true,exponent-product=\cdot]{siunitx}
\DeclareSIUnit\angstrom{\text{\AA}}

\usepackage{todonotes}

\makeatletter
\AtBeginDocument{\let\LS@rot\@undefined}
\makeatother

\begin{document}

\title{Probing the effects of broken symmetries in machine learning}

\author{Marcel F. Langer}
\affiliation{Laboratory of Computational Science and Modeling and National Centre for Computational Design and Discovery of Novel Materials MARVEL, Institute of Materials, \'Ecole Polytechnique F\'ed\'erale de Lausanne, 1015 Lausanne, Switzerland}

\author{Sergey N. Pozdnyakov}
\affiliation{Laboratory of Computational Science and Modeling and National Centre for Computational Design and Discovery of Novel Materials MARVEL, Institute of Materials, \'Ecole Polytechnique F\'ed\'erale de Lausanne, 1015 Lausanne, Switzerland}

\author{Michele Ceriotti}
\email{michele.ceriotti@epfl.ch}
\affiliation{Laboratory of Computational Science and Modeling and National Centre for Computational Design and Discovery of Novel Materials MARVEL, Institute of Materials, \'Ecole Polytechnique F\'ed\'erale de Lausanne, 1015 Lausanne, Switzerland}

\newcommand{\mc}[1]{{\color{magenta}#1}}
\newcommand{\MLi}[1]{{\color{orange}#1}}
\newcommand{\snp}[1]{{\color{purple}#1}}

\date{\today}%

\begin{abstract}
Symmetry is one of the most central concepts in physics, and it is no surprise that it has also been widely adopted as an inductive bias for machine-learning models applied to the physical sciences. 
This is especially true for models targeting the properties of matter at the atomic scale. Both established and state-of-the-art approaches, with almost no exceptions, are built to be exactly equivariant to translations, permutations, and rotations of the atoms. 
Incorporating symmetries -- rotations in particular -- constrains the model design space and implies more complicated architectures that are often also computationally demanding. There are indications that non-symmetric models can easily learn symmetries from data, and that doing so can even be beneficial for the accuracy of the model.
We put a model that obeys rotational invariance only approximately to the test, in realistic scenarios involving simulations of gas-phase, liquid, and solid water. 
We focus specifically on physical observables that are likely to be affected -- directly or indirectly -- by symmetry breaking, finding negligible consequences when the model is used in an interpolative, bulk, regime. Even for extrapolative gas-phase predictions, the model remains very stable, even though symmetry artifacts are noticeable.  
We also discuss strategies that can be used to systematically reduce the magnitude of symmetry breaking when it occurs, and assess their impact on the convergence of observables.
\end{abstract}

\maketitle

Data-driven techniques are increasingly applied across the physical sciences~\cite{carl+19rmp,degr+22nature}, with the modeling of matter at the atomic scale being a field in which they have been adopted early~\cite{das+06pnas,behl-parr07prl,bart+10prl,rupp+12prl} and with great success.
Machine-learning models that are meant to reproduce the relationship between a structure and its properties  inherit the constraints, and hence the symmetries, of the underlying physics.
For instance, the potential energy, the target of so-called \mlps, is invariant to atom label permutations, as well as translations, rotations, and reflections.
Ensuring that \mlps respect the inherent symmetries of the problem has long been considered essential~\cite{behl-parr07prl,bart+10prl,mgcc2021q,lgr2022q}.
The simplest approach to constructing invariant models is to use invariant features from the start, for instance interatomic distances or angles, that however leads to models with reduced descriptive power~\cite{pozd+20prl}.
Alternatively, models can rely on an equivariant architecture~\cite{s2021q}: Internal features are constructed to transform with the coordinate frame, and can then be combined into invariants for the final energy prediction. Most state-of-the-art \mlps, for instance \nequip~\cite{bmsk2022q}, \mace~\cite{bkoc2022q}, or \sok~\cite{fum2022q}, are based on this type of architecture. However, ensuring equivariance imposes severe constraints on model architectures~\cite{niga+22jcp2,bata+22arxiv}, and general equivariant operations can become computationally costly in practice.

For this reason, there has been growing interest in \scare{unconstrained} models that relax the requirement of global invariance (or internal equivariance), and that are used widely in computer science for tasks involving the classification of point clouds~\cite{qi+17nips,xu+21ieee,zhao+21ieee}.
Even in the field of atomic-scale modeling, recent work has shown that non-invariant models can achieve competitive accuracy on benchmark datasets when compared with invariant models, both for constructing \mlps~\cite{pozd-ceri23nips} and for tasks involving the prediction of the secondary structure of polypeptides~\cite{abra+24nature}.
However, good predictive performance on static test datasets is not sufficient to evaluate the practical usefulness of a given model architecture~\cite{fwgj2023q}.
Symmetries are associated with conservation laws that are beneficial for the numerical stability of algorithms~\cite{herb-levi21jpcm}, and whose violation can occasionally lead to manifestly absurd simulation outcomes~\cite{gong+07nn,wong+10nn}. 
This work investigates the impact of neglecting rotational invariance in \mlps, and to what extent an approximation of invariance is sufficient in practice.

We use the \pet architecture~\cite{pozd-ceri23nips} that is exactly invariant to translations and atom index permutations but \emph{not} to rigid rotations to train a \mlp for bulk water.
We use the training set from Ref.~\citenum{chen+19pnas} that contains \num{1593} configurations computed at the revPBE0~\cite{zhan-yang98prl,adam-baro99jcp} level of theory, including D3 dispersion corrections~\cite{grim+10jcp}.
The details of the model and the training protocols are discussed in the \SM, and can also be found in the data record associated with this publication. %
For the purpose of this study, it is important to stress that -- as in Ref.~\citenum{pozd-ceri23nips}, and as standard practice in fields using non-symmetric architectures -- rotational augmentation is performed during training: For each epoch, a different random orientation is chosen for every structure.
In addition, we implement an inference-time approximate symmetrization scheme, i.e., averaging predictions over multiple rotations, based on systematically-convergent grids over Euler angles~\cite{khal+16ieee}. This provides a way to assess the impact of symmetry-breaking on the model accuracy without changing its architecture.
Taking the base model $y(A)$, a rotationally averaged version is defined as:
\begin{equation}
\bar{y}(A)=\frac{1}{M} \sum_{k=1}^M w_k y(\hat{R}_k A) \,,
\end{equation}
where $\hat{R}_k$ indicate the uniformly-distributed rotation matrices and $w_k$ the associated quadrature weights, and $A$ the structure for which we are making a prediction. 
We indicate the grid size with the notation $N[\inv]$, where $N\ge 2$ is an integer that indicates the subdivision of the Euler angles, and $\inv$ indicates that the grid is duplicated to also include the corresponding improper rotations $\{-\hat{R}_k\}$. Grids labeled by $2$, $2\inv$, $3$, $3\inv$ contain 18, 36, 75, 150 rotations respectively. By increasing the grid size, the model can be made as close to exactly equivariant as desired, at correspondingly increased computational cost.
Note that Ref.~\citenum{pozd-ceri23nips} also proposes an \emph{exact} symmetrization scheme, which however would require modifications to the \pet architecture to enable an efficient application (see the \SM).

\begin{figure}[tbp]
\includegraphics[width=1.0\linewidth]{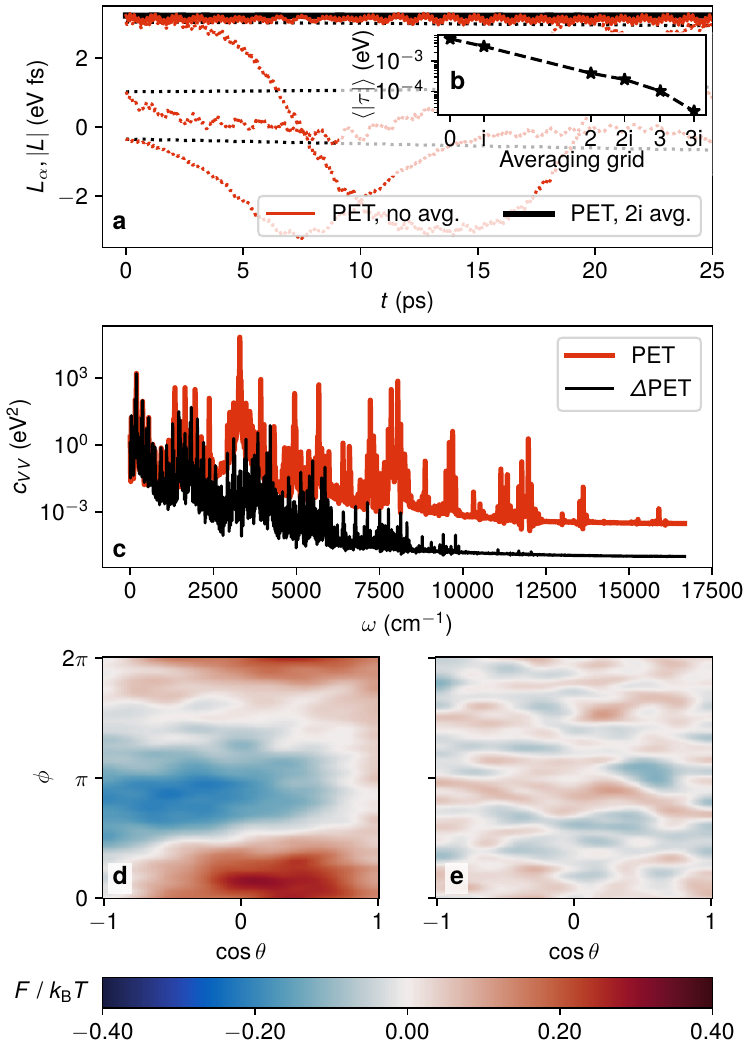}
\caption{Simulations of a water molecule using a rotationally non-equivariant \pet model. (a) Trajectories of the angular momentum components (dashed lines) and modulus (full line) during constant-energy molecular dynamics, for the model without symmetrization (red) and with rotational averaging over a $2\inv$ grid. (b) Mean value of the torque acting on the molecule over a constant-temperature simulation, for different orientation grids (using the notation $N[\inv]$). (c) Power spectrum computed from the autocorrelation function of the potential energy, and on the non-equivariant part of the potential $\Delta$ (computed as the difference between the raw model and a $2\inv$ average). (d-e) Orientational free energy for the water molecule computed over a long constant-temperature simulation without (d) and with $2\inv$ rotational averaging (e). 
\label{fig:gas-lines}}
\end{figure}

As a first, and perhaps the most extreme, test, we run constant-energy \md simulations for a water molecule in vacuum.
Given that the model is trained exclusively on bulk structures, this amounts to a deep extrapolative regime, and allows us to test the most direct consequence of the lack of rotational invariance -- break-down of angular momentum conservation.
A first observation is that the potential is very stable despite the extrapolative conditions, and can be run for several nanoseconds with energy conservation consistent with the time step of \SI{0.5}{fs} and the use of single-precision arythmetics.
The symmetry breaking is however apparent in the precession of the angular momentum $\mathbf{L}$ (Fig.~\ref{fig:gas-lines}a) that is a consequence of the non-zero torque acting on the molecule despite the absence of an external potential (Fig.~\ref{fig:gas-lines}b). 
The torque $\boldsymbol{\tau}$ is almost orthogonal to the angular momentum, so the angular velocity is almost constant. The small fluctuations of the total momentum are an indication of the coupling of the non-equivariant terms with the internal degrees of freedom of the molecule. 
Rotational averaging mitigates symmetry breaking, and systematically reduces the magnitude of $|\boldsymbol{\tau}|$ -- which does not eliminate precession, but slows it down dramatically, and effectively eliminates the fluctuations of $|\mathbf{L}|$. 

Another important observation is that the non-equivariant component of the potential (estimated as the difference between the single and rotationally averaged predictions of the model for each structure) shows fluctuations that are not only much smaller than those of the actual potential, but also slowly-varying (Fig.~\ref{fig:gas-lines}c).
This means it is possible to apply multiple time-step (MTS) methods~\cite{tuck+92jcp} and avoid evaluating the averaged model, which is computationally more demanding, at every \md step. In all of the constant-temperature simulations performed in this work that use rotational averaging, we use the MTS implementation in i-PI~\cite{kapi+19cpc}, with an inner time step of \SI{0.5}{fs} for the base model and evaluating the rotationally averaged forces every \num{10} steps. 

Angular momentum precession for an isolated system is a telltale sign of $\mathrm{SO}(3)$ symmetry breaking, but precise classical dynamics is only relevant in few molecular applications, such as the study of gas-phase chemical reactions~\cite{farr-lee74arpc,mikl01cpc}. 
A much more common scenario involves simulations that sample a thermal distribution and compute statistical averages over the trajectory.
In this case, a clear signature of broken symmetry would be a preferential absolute orientation of the water molecules in space. We assess this by computing a histogram of the polar angle of the molecular orientation (defined as the vector connecting the oxygen atom with the mid-point of the two hydrogen atoms), over a long trajectory that is supplemented with an efficient colored-noise thermostat~\cite{ceri+10jctc} to sample a classical Boltzmann distribution at $T=300$~K. 
The histogram can then be re-cast as a free energy that should be constant throughout the spherical coordinate system. 
As shown in Fig. \ref{fig:gas-lines}d there is indeed a significant (but \emph{tiny}) inhomogeneity, of the order of a fraction of $k_{\mathrm{B}} T$. 
Rotational averaging brings the anisotropy down to the level of statistical noise (Fig. \ref{fig:gas-lines}e), which is consistent with the sharp reduction in the value of the torque seen in Fig.~\ref{fig:gas-lines}b.  

\begin{figure}[tbp]
    \centering
    \includegraphics[width=1.0\linewidth]{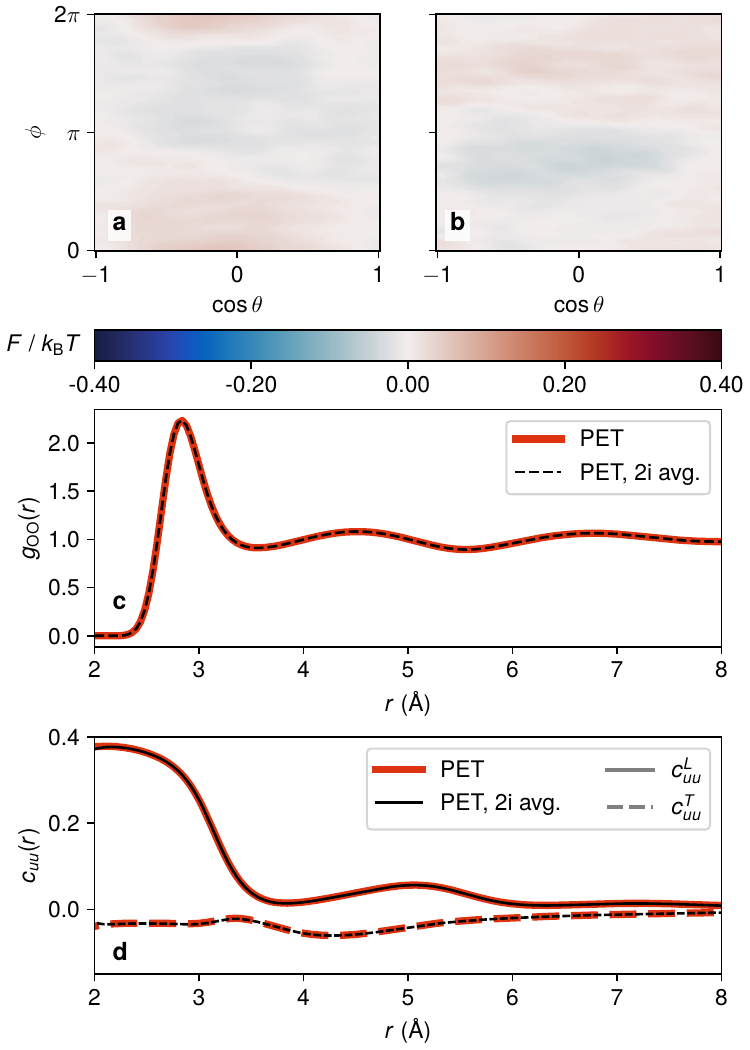}
    \caption{Structural properties of liquid water at $T=300$~K, simulated with a \pet model with and without $2\inv$ rotational averaging. 
    (a-b) Orientational free energy for the water molecule computed over a long constant-temperature simulation without (a) and with (b) $2\inv$ averaging .
    (c) O-O pair correlation function. (d) Molecular orientation correlation function, computed separately for the longitudinal (full lines) and transverse (dashed lines) components.}
    \label{fig:bulk-static}
\end{figure}

\begin{figure}[tbp]
    \centering
\includegraphics[width=1.0\linewidth]{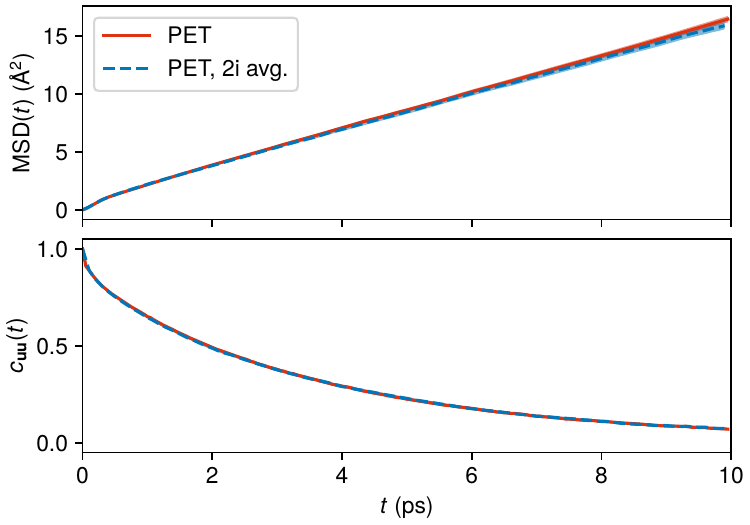}
    \caption{Dynamical properties of liquid water at $T=300$~K, simulated with a \pet model with and without a $2\inv$ rotational averaging.
    The shaded area around the curves indicates the (small) statistical uncertainty. 
    (a) Oxygen mean-square displacement curves, whose slope is proportional to the diffusion coefficient. (b) Dipole autocorrelation function, which is indicative of the rotational dynamics of water molecules.}
    \label{fig:bulk-dyn}
\end{figure}

We now move to the more typical use case of simulations of bulk water, to assess whether the small, but measurable, violation of isotropy for the isolated molecule has a more significant impact on the collective behavior of matter in the condensed phase. 
We run a long classical MD trajectory (1~ns) of a relatively large box (512 molecules) at room temperature, in the $NVT$ ensemble at $T=300$~K, using a stochastic velocity rescaling thermostat~\cite{buss+07jcp} that has a negligible effect on dynamical properties~\cite{buss-parr08cpc}.
Given that periodic boundary conditions make it impossible to define and monitor a conserved angular momentum, we look for a signature of non-equivariant behavior in the  absolute orientation of the water molecules. 
The free energy profile is almost perfectly isotropic even without rotational averaging (Fig.~\ref{fig:bulk-static}a-b), indicating that not only there are no collective effects that generate spurious molecular orientations, but that for thermodynamic conditions that are well-represented in the training data, the \pet model is even closer to being exactly equivariant. 
We can further assess indirect effects of the small symmetry breaking by computing structural properties of the liquid, such as the pair correlation function and dipole-dipole correlations. These quantities, which depend subtly on the relative position and orientation of pairs of water molecules, are essentially left unchanged by the application of inference-time averaging (Fig.~\ref{fig:bulk-static}c-d) -- indicating that the lack of exact equivariance in the raw \pet predictions is inconsequential.
Even though it appears that structural properties of water are perfectly converged without rotational averaging, one may wonder if \emph{dynamical} properties, which are strongly dependent on the height of energy barriers, would be more sensitive to the broken rotational symmetry. 
Fig.~\ref{fig:bulk-dyn} shows that this is not the case: Both translational and orientational diffusion are identical within the statistical error, regardless of whether the \pet model is made more equivariant by averaging over a grid of rotations. 

As a final test, we consider the energetics of proton disorder in hexagonal ice. We consider 9 proton-disordered cells from Ref.~\citenum{hayw-reim97jcp}, and optimize the geometry using a raw \pet model and a $2\inv$ rotational average. 
This amounts to an intermediate degree of extrapolation: Even though the training set contains only disordered structures, it has been shown that models trained on liquid water are also capable of describing, with good accuracy, the solid portion of the phase diagram~\cite{mons+20nc}.
It is also a problem for which small energy differences matter and a case in which a small preference for a particular orientation could easily lead to macroscopic distortions upon relaxation.
Once again, the practical impact of approximate equivariance is negligible. The forces on the initial structures (that are of the order of \SI{1}{eV\per\angstrom}) differ by less than \SI{1}{meV\per\angstrom} between standard and rotationally averaged \pet. 
Even though individual proton-ordered structures have energies that differ from each other by only \SI{0.3}{meV\per molecule}\footnote{Note that even though this is a stringent test for rotational symmetry breaking, the energies are unlikely to fully capture the physics of proton ordering, given that \pet, as most \mlps, is a local model and misses an explicit description of long-range electrostatics.}, relative energies are predicted by the base model with an error that is an order of magnitude smaller, about \SI{0.02}{meV\per molecule}. The relaxed geometries have a minuscule \gls{rmsd} of about \SI{0.001}{\angstrom\per atom}.

Our tests show that applying random rotations during training, i.e., standard data augmentation, can be sufficient to achieve a very high degree of approximate equivariance. 
There are essentially no measurable effects on the static and dynamical properties obtained in the interpolative regime; the potential remains stable and very close to equivariant even when extrapolating to a completely different thermodynamic state point, from bulk water to a single gas-phase molecule.  
We suggest that rotational averaging during inference (either using a regular grid as we do here, or with exact symmetrization techniques that can restore rigorous equivariance~\cite{pozd-ceri23nips,nadav2024arxiv}) can be used as a safeguard and a sanity check. The associated overhead can be reduced by using a multiple-time-step integrator, or by only computing the symmetrized potential occasionally to monitor the discrepancy with the non-symmetric model.
Furthermore, there are several strategies one could apply to obtain an equivariant description avoiding this inference-time overhead entirely.
For example, one can apply a random rotation before each \pet evaluation, so that the potential is \emph{on average} independent of the absolute orientation. 
This introduces a (small) noise that disrupts energy conservation, but can be controlled with gentle thermostatting -- a strategy that is used routinely in atomistic simulations to control errors due to incomplete convergence of self-consistent algorithms~\cite{kuhn+07prl} or sampling errors in quantum Monte Carlo~\cite{mazz-sore17prl}.
This process of random rotations leads to simulations of bulk water that are free of preferential orientation effects. The small level of noise on the force is perfectly compensated by a mild, dynamics-preserving, stochastic velocity rescaling thermostat (see the \SM). 
Another possibility that would be relevant where obtaining a high level of equivariance is more important than the sheer accuracy of the energy and force predictions, is to modify the training loss to explicitly penalize symmetry breaking, e.g., evaluating the same structure over multiple orientations and requiring each prediction to match the rotational average. 
This can help pushing the degree of equivariance below the residual regression error and can also be done for out-of-sample structures for which reference properties have not been computed, serving as a form of regularization~\cite{alle+21mlst}.

Obviously, our observations are specific to the \pet architecture and the systems we considered, but they add to a growing body of empirical evidence indicating that the practical impact of neglecting rotational symmetry is usually small.
The success of non-equivariant models in other applications of geometric deep learning and computer vision~\cite{qi+17nips,guo+21cvm} is a clear example, as well as the minute effects resulting from the application of an exact symmetrization scheme on validation errors in a previous study of the \pet architecture~\cite{pozd-ceri23nips}.
We think that the fact that rotations form a compact group with a low dimension contributes to the ease by which they can be learned from relatively small data sets.
One should however keep in mind that no amount of testing can guarantee that there are no corner cases, or adversarial examples, in which a broken-symmetry model would lead to grossly unphysical predictions. 
The angular momentum precession of the isolated water molecule is a clear -- although perhaps contrived -- example.

Still, this study provides some confidence to computational physicists investigating promising non-equivariant architectures, and demonstrates simple schemes to monitor and improve the compliance with symmetry constraints at little to no cost. 
Despite the unquestionable appeal of incorporating fundamental physical concepts in the architecture of machine-learning models, it might be beneficial -- and it certainly is not as detrimental one would expect -- to just let models learn.

\begin{acknowledgments}

\section*{Supporting material}

The \pet code is freely available on \url{https://github.com/spozdn/pet/}. 
Templates for the different tests we present, and weights for the trained \pet models, will be made available upon publication.  
Additional information can also be found at \url{https://marcel.science/eqt}.

\section*{Acknowledgements}
\begin{acknowledgments}
ML and MC acknowledge funding from the European Research Council (ERC) under the European Union’s Horizon 2020 research and innovation programme Grant No. 101001890-FIAMMA. SP and MC acknowledge support from the NCCR MARVEL, funded by the Swiss National Science Foundation (SNSF, grant number 182892) and from the Swiss Platform for Advanced Scientific Computing (PASC).
\end{acknowledgments}

\end{acknowledgments}

\newcounter{sipage}
\setcounter{sipage}{1}
\loop
{%
\clearpage


\section*{Supplementary material}

# Probing the effects of broken symmetries in machine learning
## Supporting Materials

Marcel F. Langer,[1] Sergey N. Pozdnyakov,[1] and Michele Ceriotti[1, *]

[1]*Laboratory of Computational Science and Modeling and National Centre for Computational Design and Discovery of Novel Materials MARVEL, Institute of Materials, École Polytechnique Fédérale de Lausanne, 1015 Lausanne, Switzerland*

## I. POINT EDGE TRANSFORMER

The architecture we employ, Point Edge Transformer (PET), was introduced and described in detail in Ref. 1. While PET was thoroughly discussed in that cited work, we provide a summary here, with a primary focus on the hyperparameters that were modified relative to a similar training exercise for the water dataset from Ref. 2. PET is a graph neural network featuring $N_{\text{GNN}}$ message-passing layers. At each layer, messages are exchanged between all atoms within a distance $R_c$ from each other. The functional form of each layer is an arbitrarily deep transformer applied individually to each atom. Atomic environments are constructed around each atom, defined by all neighbors within $R_c$. Each neighbor sends a message to the central atom, with each message being a token of fixed size $d_{\text{PET}}$. These tokens are processed by a transformer, which performs a permutationally equivariant sequence-to-sequence transformation. The output sequence is then treated as outbound messages from the central atom to all neighbors. Consequently, for a model with $N_{\text{GNN}}$ layers and a system with $N$ atoms, there are $N_{\text{GNN}}$ individual transformers with distinct weights, each independently invoked $N$ times, resulting in $N_{\text{GNN}}N$ transformer runs. The number of input tokens for each transformer run is determined by the number of neighbors of the central atom.

In addition to an input message from a neighboring atom, geometric information about the displacement vector $\mathbf{r}_{ij}$ from the central atom to the corresponding neighbor is incorporated into the token. After each message-passing layer, all output messages are fed into a head (individual for each message-passing layer), implemented as a shallow MLP, to produce a contribution to the total prediction. The total prediction, in this case, the potential energy of the system, is computed as the sum of all head outputs over all message-passing layers and all messages. This architecture is rigorously invariant with respect to translations because it uses displacement vectors that do not change if both the central atom and a neighbor are rigidly shifted. It is invariant with respect to permutations of identical atoms because the transformer defines a permutationally covariant sequence-to-sequence transformation, and the sum over the contributions from all edges yields an overall invariant energy prediction. However, it is not rotationally invariant since it operates with the raw Cartesian components of displacement vectors.

The fitting scheme we used is identical to the one described in Ref. 1, with more details available in Appendix C.5 of the cited work. We use Adam with linear warmup and do not apply weight decay. A specific aspect we

| Model | $E$ (meV) | $F$ (meV/Å) |
|---|---:|---:|
| Cheng et al. [2], RMSE | 4500 | 120 |
| NEQUIP [3], MAE | 120 | 21 |
| PET [1], MAE | - | 14 |
| PET [this work], MAE | 74 | 17 |
| PET [this work], RMSE | 167 | 60 |

Table I. Energy and force errors for the test set for the bulk water dataset from Ref. 2, comparing the model used here with those in the literature. Note that different results in the literature report either mean absolute errors or root mean square errors, as indicated, and that we could not use the exact same test/train split (although the train fraction is consistent) so the comparison is only indicative. The observed drastic difference between MAE and RMSE indicates the presence of outliers in the dataset, with the RMSE metric being heavily influenced by them. Additionally, it is important to note that while the difference in MAE metrics for the validation (used for early stopping) and test subsets was negligible, the RMSE for PET on the validation subset was about 40 meV/Å, or about 1.5 times smaller than on the test subset. Given that the training, validation, and test split were random, this discrepancy further suggests that the RMSE metric is dominated by a few outliers.

---

[*] michele.ceriotti@epfl.ch

found beneficial for the final accuracy is fitting the model with a high learning rate initially and then decreasing the learning rate very rapidly at a certain point. Thus, our learning rate scheduler loosely resembles but is not identical to OneCycleLR [4]. The hyperparameters of the architecture we used differ slightly from those in Ref. 1 to make the model noticeably faster at the cost of a slight deterioration in accuracy. Similar to Ref. 1, $d_{\text{PET}}$ was set to 128, and $N_{\text{TL}}$ (number of self-attention layers in each transformer) to 2. The most accurate model in Ref. 1 for the discussed dataset used $R_c = 4.25$ Å and $N_{\text{GNN}} = 6$. In contrast, we used $R_c = 4.0$ Å and $N_{\text{GNN}} = 4$. Consequently, the accuracy slightly drops from a mean absolute error (MAE) on force components of 14.4 meV/Å, as reported in Ref. 1, to 17.1 meV/Å. However, the model becomes faster, making our selection of hyperparameters a reasonable trade-off between accuracy and computational efficiency.

We report the validation error in comparison with some results in the literature (Table I), with the sole purpose of demonstrating that the model we use is competitive with state-of-the-art equivariant models from the point of view of benchmark accuracy.

## II. APPROXIMATE (AND EXACT) ROTATIONAL SYMMETRIZATION.

Throughout our experiments, we used an approximate symmetrization scheme defined in the main text, which involves averaging predictions over multiple rotations based on systematically convergent grids over Euler angles. This approach allows a model to be made arbitrarily close to an equivariant one, albeit at the cost of progressively increased computational demand. However, it still does not achieve rigorous equivariance. Given that Ref. 1 proposes an *exact* symmetrization scheme, termed the Equivariant Coordinate System Ensemble (ECSE), it is worth explaining why we decided not to employ it in this work. The exact symmetrization protocol works by defining a set of local coordinate systems (equivalently rotations) for each atomic environment that are *rigidly attached* to it and, thus, rotate synchronously with the atomic environment. For strictly local models, the final prediction is computed by running a backbone architecture for each of these rotations and then computing a weighted sum of all the predictions. Since all the coordinate systems rotate synchronously with the atomic environment, all the predictions of the backbone architecture are exactly invariant with respect to rotations, ensuring the final prediction is also rotationally invariant. The use of an ensemble of local coordinate systems, rather than selecting an individual frame, aims to ensure smoothness of the resulting symmetrized model with respect to geometric deformations of an input atomistic system, which is challenging to achieve using only one local coordinate system [5]. The ECSE scheme also include a several optimizations to reduce the number of coordinate systems needed to achieve smooth averaging, which makes ECSE computationally efficient for strictly local models.

For message-passing schemes, and PET in particular, such an approach encounters a difficulty: when a message is sent from atom A to atom B, there is a mismatch between the coordinate systems defined for the atomic environments around atom A and atom B. Thus, for message-passing schemes, one possibility is to utilize a naive approach by explicitly treating message-passing schemes as local models (where the cutoff radius is their receptive field). While this maintains linear scaling, it is very computationally expensive. Alternatively, one would need to redesign the message-passing mechanism by allowing one to "match" the coordinate systems of the different local environments, which changes the nature of the model and is therefore incompatible with our goal to compare the raw predictions of a non-symmetrized PET model with one that has been made (more) equivariant without changing its architecture.

### III. ROTATIONAL ERROR FOR THE POTENTIAL

We can assess the magnitude of the symmetry breaking in a very direct way by comparing the raw prediction of the energy and forces of the PET model with those computed with a high degree of rotational averaging. As shown in Table II, the error on forces an energy for the test set is about 10 times smaller than the error relative to the DFT reference. Comparing the energy error for structures obtained from NVT trajectories with that on the test set is not trivial, as the system-size scaling depends on whether the errors on atom and bond-centered contributions are systematic, or uncorrelated. However, it is clear from the force errors – that are intensive and therefore easier to compare between structures of different size – that the bulk trajectories are in an interpolative regime, with smaller errors than for the test set (which contains highly-distorted configurations from high-temperature simulations) while the isolated molecule have a much larger symmetry error, consistent with the extrapolative nature of the prediction. Still, the rotational error for the energy of the molecule is one order of magnitude smaller than thermal energy at the simulated conditions, which is reflected in the very weak anisotropy of the orientational free energy. Fig. 1 demonstrates the convergence of the force error with increasing degree of rotational averaging, demonstrating that the 2i grid reduces, in all cases, the rotational error by at least an order of magnitude.

| Dataset | $\langle|E_{\mathrm{PET}} - E_{3\mathrm{i}}|\rangle$ (meV) | $\langle|F_{\mathrm{PET}} - F_{3\mathrm{i}}|\rangle$ (meV/Å) |
| --- | --- | --- |
| Test set ($n_{\mathrm{H_2O}} = 64$) | 5.1 | 2.0 |
| Bulk MD ($n_{\mathrm{H_2O}} = 512$) | 8.0 | 1.2 |
| Gas MD ($n_{\mathrm{H_2O}} = 1$) | 2.7 | 12.0 |

Table II. Symmetry error (MAE) for total energy and forces, computed over the test set, a collection of snapshots from a NVT bulk simulation at 300 K, and a collection of snapshots from a NVT gas-phase simulation at 300 K.

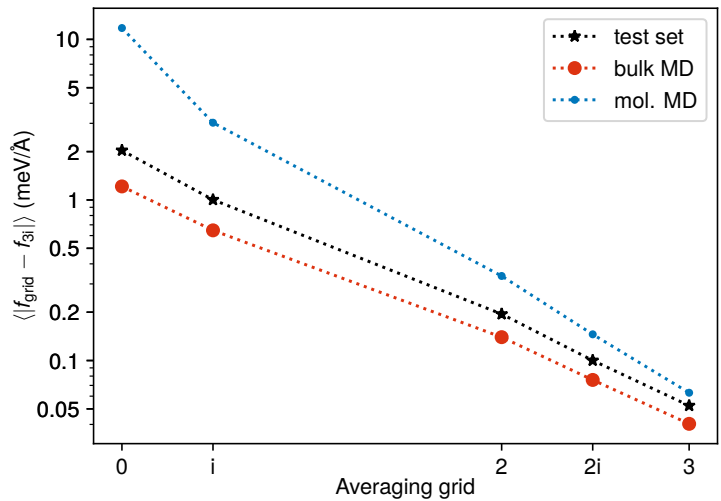

Figure 1. Convergence of rotational error for the forces (MAE) computed over the test set, a collection of snapshots from a NVT bulk simulation at 300 K, and a collection of snapshots from a NVT gas-phase simulation at 300 K. The grids are the same discussed in the main text.

## IV. STABILITY OF THE DYNAMICS

The PET model yields excellent qualitative stability – meaning that trajectories of several nanoseconds can be performed for both bulk and gas-phase water without observing catastrophic events in which molecules lose their chemical integrity. More quantitatively, Fig. 2 shows a small drift of the conserved quantity – defined as potential plus kinetic energy, and including an additional term that tracks the total heat balance of the thermostat used to enforce constant-temperature sampling[6] for the trajectories using non-symmetrized PET. This very small drift is not due to the lack of rotational equivariance (as PET is still rigorously conservative) but to the fact we use single-precision arithmetics. The drift is larger – but still amounting to a few tens of meV per molecule per nanosecond – when using a multiple time step integrator with the grid-averaged PET evaluated every ten $0.5\,\mathrm{ps}$ steps, and when performing on-the-fly averaging by using a different random rotation at each evaluation of the model. Even in this latter case, a global stochastic velocity rescaling thermostat is sufficient to maintain accurate canonical sampling: Structural properties are indistinguishable from those of the grid-averaged trajectory (as shown in the next Section) and the mean kinetic energy of O and H atoms is less than $0.2\,\mathrm{K}$ away from the target temperature of $300\,\mathrm{K}$.

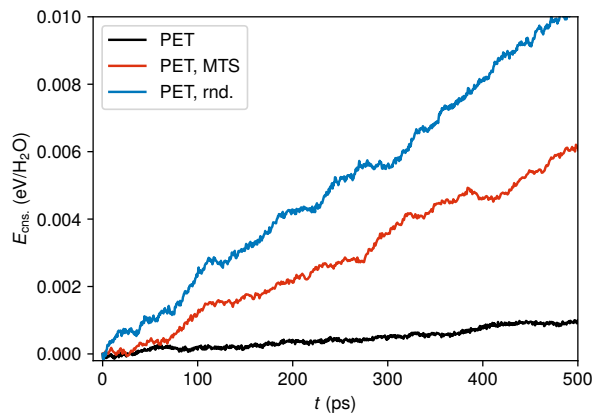

Figure 2. Drift of the conserved quantity (total energy plus heat balance of the thermostat) for simulations of liquid water using a non-symmetrized PET model, a $2i$-grid rotation averaging with multiple time step integration (MTS), and an on-the-fly random augmentation protocol (rnd.).

## V. ON-THE-FLY RANDOM AUGMENTATION

We perform simulations for bulk water using at each force evaluation a different random rotation of the system. We use a global stochastic velocity rescaling thermostat [7] that does not affect significantly the dynamical properties of the system to control the effect of the resulting noise, which leads to a noticeable, but small, increase in the drift of the conserved quantity, as observed in Fig. 2. Figures 3 and 4, to be compared with the corresponding plots in the main text, demonstrate that the orientational distribution of the water molecules is isotropic within the statistical noise, and that all static and dynamic translational and rotational correlations are indistinguishable from that of a run using a grid-averaged model.

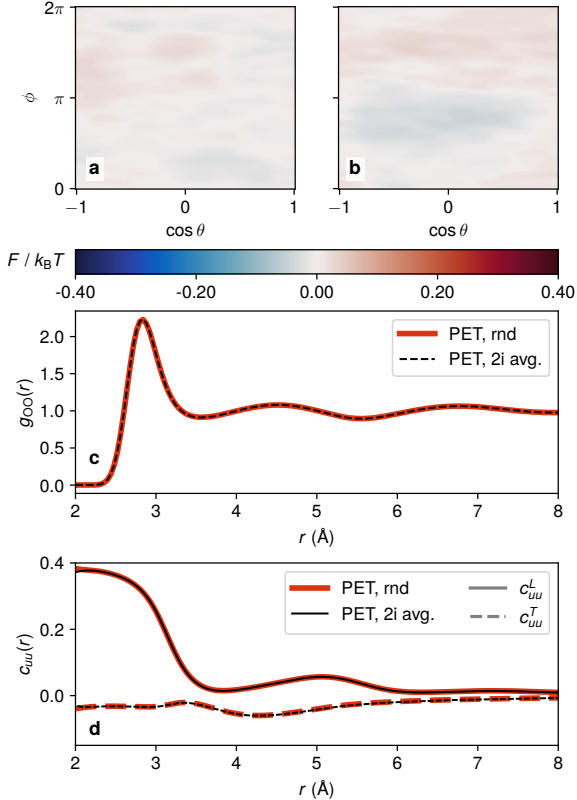

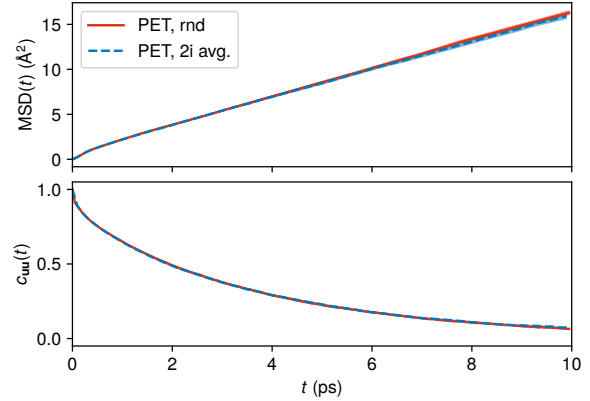

Figure 4. Dynamical properties of liquid water at $T = 300$ K, simulated with a PET model with random and 2i-grid rotational averaging. (a) Oxygen mean-square displacement curves, whose slope is proportional to the diffusion coefficient. (b) Dipole autocorrelation function, which is indicative of the rotational dynamics of water molecules.

Figure 3. Structural properties of liquid water at $T = 300$ K, simulated with a PET model with on-the-fly random augmentation, and with a fixed 2i grid rotational averaging. (a-b) Orientational free energy for the water molecule computed over a long constant-temperature simulation with random (a) and $2i$ (b) averaging. (c) O-O pair correlation function. (d) Molecular orientation correlation function, computed separately for the longitudinal (full lines) and transverse (dashed lines) components.


}
\addtocounter{sipage}{1}
\ifnum \value{sipage}<7
\repeat

\end{document}